\documentstyle[preprint,eqsecnum,aps]{revtex}
\begin{document}

\draft

\title{Dynamics of Interfaces in Superconductors}

\author{Alan T. Dorsey}

\address{Department of Physics, University of Virginia\\
McCormick Road, Charlottesville, VA 22901}


\maketitle

\begin{abstract}
The dynamics of an interface between the normal and superconducting phases
under nonstationary external conditions is studied within the framework
of the time-dependent Ginzburg-Landau equations of superconductivity,
modified to include thermal fluctuations.  An equation of
motion for the interface is derived in two steps.  First, the
method of matched asymptotic expansions is used to derive a diffusion
equation for the magnetic field in the normal phase, with nonlinear
boundary conditions at the interface.  These boundary conditions
are a continuity equation which relates the gradient of the field at the
interface to the normal velocity of the interface, and a
modified Gibbs-Thomson boundary condition for the  field at the interface.
Second, the boundary integral method is used to integrate out the
magnetic field in favor of an equation of motion for the interface.
This equation of motion, which is highly nonlinear and nonlocal, exhibits
a diffusive instability (the Mullins-Sekerka instability)
when the superconducting phase  expands into the
normal phase (i.e., when the external field is reduced below the
critical field).   In the limit of infinite diffusion constant
the equation of motion
becomes local in time, and can be derived variationally from a static
energy functional which includes the bulk free energy
difference between the two phases, the interfacial energy, and a long
range self-interaction of the interface of the Biot-Savart form.
In this limit the  dynamics is  identical
to the interfacial dynamics of ferrofluid domains recently proposed
by S. A. Langer {\it et al}. [Phys. Rev. A {\bf 46}, 4894 (1992)].
As shown by these authors, the Biot-Savart interaction leads to mechanical
instabilities of the interface, resulting in highly branched labyrinthine
patterns.  The application of these ideas to the study of
labyrinthine patterns in the intermediate state of type-I superconductors
is briefly discussed.

\end{abstract}


\section{Introduction}

Over the last decade there has evolved a fairly complete understanding of
the physics of several types of nonequilibrium growth patterns, such
as the dendritic growth of solidifying systems (e.g., ``snowflakes'')
or the fingered growth which occurs at the interface of driven
immiscible fluids (e.g., viscous fingers in Hele-Shaw cells);
for reviews of the theoretical situation, see Kessler {\it et al}.\
\cite{kessler88} and Langer \cite{langer88}.  The similar patterns which
grow in these ostensibly different physical systems are the consequence
of a competition between a dynamic instability (the Mullins-Sekerka
instability \cite{mullins64} for dendritic growth) which promotes the growth of
a highly
ramified interface, and surface tension, which favors a smooth interface.
One of the important theoretical insights which has emerged from this work
is that surface tension anisotropy
plays a crucial role in determining the morphology of the pattern
\cite{kessler88,langer88}.

It has recently been shown that the process of magnetic flux expulsion
in type-I superconductors subjected to a magnetic field quench share
many features with these other pattern forming systems \cite{frahm91,liu91}.
In particular, as a superconducting nucleus grows, the expelled flux
generates eddy currents in the normal phase; the magnetic field ${\bf h}$
in the normal phase therefore satisfies a diffusion equation,
\begin{equation}
\partial_{t} {\bf h} = D \nabla^{2} {\bf h},
\label{intro1}
\end{equation}
where the magnetic diffusion constant $D=1/4\pi\sigma$, with
$\sigma$ the normal state conductivity (we will set $c=1$).
By applying Maxwell's equations to
the interface itself, and noting that the electric and magnetic fields
both vanish in the superconducting phase, we arrive at a continuity equation
for the field at the interface \cite{pippard50,lifshitz50,andreev68},
\begin{equation}
(\nabla\times {\bf h})\times\hat{\bf n}|_{i} = -  D v_{n} {\bf h}_{i},
\qquad {\bf h}\cdot \hat{\bf n} |_{i} = 0,
\label{intro2}
\end{equation}
where $\hat{\bf n}$ is the unit normal at the interface, directed toward
the normal phase, and $v_{n}$ is the normal velocity at the interface.
Finally, the field at the interface should equal the superconducting
critical field $H_{c}$, with curvature corrections:
\begin{equation}
|{\bf h}|_{i} = H_{c} \left\{ 1 - {4\pi\over H_{c}^{2}}\left[
  \sigma_{\rm ns}(\theta) + \sigma_{\rm ns}''(\theta)\right]{\cal K} \right\},
\label{intro3}
\end{equation}
where  $\sigma_{\rm ns}(\theta)$ is the surface tension of the
normal/superconducting interface (not to be confused with the conductivity),
which depends on the angle $\theta$ with respect to the crystal axes
(it would depend on two angles $\theta_{1,2}$ in three dimensions),
and ${\cal K}$ is the curvature of the interface (or the sum of the
principle curvatures in  three dimensions) \cite{kuper51,frahm91,liu91}.
Similar equations (without the surface tension)
were used by Pippard \cite{pippard50} and Lifshitz \cite{lifshitz50}
to discuss the growth of the normal phase into the superconducting
phase, which is a dynamically stable process.
The analogy with the solidification
problem is apparent when the magnetic field is identified with the
temperature of the liquid, and the magnetic diffusion constant with the
thermal diffusivity;  the continuity condition, Eq. (\ref{intro2}),
is replaced with the continuity condition for the heat generated at the
solidifying interface, and Eq.\ (\ref{intro3}) is a modified form of the
Gibbs-Thomson boundary condition \cite{kessler88,langer88}.   Due to this
formal similarity
between the equations describing solidification and those describing the
kinetics of the normal/superconducting transition, it was predicted that
the growth of a superconducting nucleus into the ``supercooled'' normal
state would be dynamically unstable \cite{frahm91,liu91}.  Such
instabilities were observed in numerical studies of the time-dependent
Ginzburg-Landau (TDGL) equations of superconductivity for propagating
interfaces; see e.g., Fig. 1 of Ref.\ \cite{frahm91}.

In this paper we will connect the TDGL equations to the ``sharp-interface''
equations discussed above.  In Sec.\ II the TDGL equations, including
fluctuations, are presented.  In Sec.\ III we use the method of
matched asymptotic expansions to reduce these equations
to a set of sharp interface equations, generalized to include fluctuations.
In Sec.\ IV we formally reduce the sharp interface equations to an
equation of motion for the interface itself, and discuss some of the
features of this equation of motion.  In particular, we will show that in the
limit that the diffusion constant $D\rightarrow\infty$, the interface
equation of motion becomes local, and can be derived from an
interfacial energy functional.  This
functional contains three terms: the bulk free energy difference
between the normal and superconducting phases, the interfacial surface
energy, and a Biot-Savart interaction of the interface with itself.
When written in this form, the equation of motion is identical to
a phenomenological equation of motion for the dynamics of two dimensional
domains of ferrofluids,  recently proposed by
Langer {\it et al}. \cite{langer92}.  Appendix~A reviews some properties
of the surface tension, along with a new result on the behavior of the
surface tension near the critical value of $\kappa=1/\sqrt{2}$.
Appendix~B includes some details on the calculation of the kinetic
coefficient which appears in the equation of motion.  In Appendix~C
the noise correlations are calculated and shown to satisfy the
fluctuation-dissipation theorem.  Some useful definitions and results
concerning the differential geometry of curves in two dimensions are
provided in Appendix~D.

\section{The time-dependent Ginzburg-Landau model}

The time-dependent Ginzburg-Landau (TDGL) model consists of equations
of motion for the complex superconducting order parameter $\psi$,
the magnetic vector potential ${\bf a}$, and the scalar potential
$\phi$ \cite{schmid66}.  The origin and validity of the equations have been
extensively discussed elsewhere \cite{gorkov75,dorsey92}.
In conventional units, these equations are
\begin{equation}
\hbar\gamma (\partial_{t} + i {e^{*}\over \hbar} \phi) \psi =
{\hbar^{2} \over 2 m} (\nabla - i {e^{*}\over \hbar} {\bf a} )^{2} \psi
+ |a| \psi - b|\psi|^{2} \psi + \theta,
\label{tdgl1}
\end{equation}
\begin{equation}
\nabla\times\nabla\times {\bf a} = 4\pi ( {\bf J}_{n} + {\bf J}_{s}
  + \tilde{\bf J} ),
\label{tdgl2}
\end{equation}
where the normal current ${\bf J}_{n}$ is given by
\begin{equation}
{\bf J}_{n} = \sigma {\bf e} = \sigma (-\nabla \phi - \partial_{t} {\bf a}),
\label{tdgl3}
\end{equation}
and where the supercurrent ${\bf J}_{s}$ is given by
\begin{equation}
{\bf J}_{s} = {\hbar e^{*} \over 2 m i}(\psi^{*}\nabla\psi
   - \psi\nabla\psi^{*}) - {(e^{*})^{2}\over m} |\psi|^{2} {\bf a}.
\label{tdgl4}
\end{equation}
In these equations $\gamma$ is a dimensionless order parameter relaxation
time; $e^{*}$ and $m$ are the charge and mass of a Cooper pair;
$|a| = a_{0}(1 - T/T_{c0})$, with $T_{c0}$ the mean-field transition
temperature; $\sigma$ is the conductivity of the normal phase.  We have
also included two noise terms, $\theta$ and $\tilde{\bf J}$, which
are choosen so as to generate the correct Boltzmann weights in
equilibrium \cite{hohenberg77}.   We then have
$\langle \theta \rangle = \langle \tilde{\bf J} \rangle = 0 $, and
\begin{equation}
\langle \theta^{*} ({\bf x},t)\theta ({\bf x'},t')\rangle = 2 \hbar\gamma
k_{B} T \delta^{(d)}({\bf x} - {\bf x}') \delta(t-t'),
\label{noise1}
\end{equation}
\begin{equation}
\langle \tilde{J}_{i}({\bf x},t)\tilde{J}_{j}({\bf x}',t')\rangle
= 2 \sigma k_{B} T \delta^{(d)}({\bf x} - {\bf x}') \delta(t-t') \delta_{ij},
\label{noise2}
\end{equation}
with all cross correlations zero (the brackets
denote an average with respect to the noise distribution).  In terms of the
parameters in the TDGL equations, the correlation length
$\xi=\hbar/(2 m |a|)^{1/2}$, the penetration depth
$\lambda = [mb/4\pi (e^{*})^{2}|a|]^{1/2}$, the Ginzburg-Landau parameter
$\kappa = \lambda/\xi$, and the thermodynamic critical
field $H_{c} = (4\pi |a|^{2}/b)^{1/2}$.

In deriving a set of sharp interface equations it will be useful to work
with a judiciously chosen set of dimensionless variables.
Sharp interfaces  between the superconducting and normal phases will be
produced when the coherence length $\xi$ is small; i.e., when $|a|$ is
large (this implies that we are far from the normal/superconducting
phase boundary).  We will therefore introduce a small parameter $\epsilon$
such that $|a|=\bar{a}/\epsilon^{2}$, with $\bar{a}$ a fixed constant.
To recast the TDGL equations into dimensionless form we introduce the
following primed dimensionless variables,
\begin{equation}
\begin{array}{lll}
{\bf x} =\bar{ \lambda} {\bf x}',&  t = (\hbar\gamma/\bar{a})t',
   & \psi = (|a|/b)^{1/2} \psi' \nonumber\\
{\bf a} = \sqrt{2}\bar{ H}_{c}\bar{\lambda} {\bf a}',
       &  \phi = (\bar{a}/e^{*}\gamma)\phi' & ,
\label{dimension1}
\end{array}
\end{equation}
along with dimensionless conductivity and temperature variables
\begin{equation}
\begin{array}{ll}
\bar{\sigma} = 4\pi\kappa^{2} (\hbar/2m \gamma) \sigma, &
\qquad \bar{T} = k_{B} T /[(\bar{H}_{c}^{2}/4\pi)\bar{ \lambda}^{d}].
\label{dimension2}
\end{array}
\end{equation}
Here we have defined $\bar{\lambda}= [mb/4\pi (e^{*})^{2}\bar{a}]^{1/2}$,
$\bar{\xi} = \hbar/(2m\bar{a})^{1/2}$, and
$\bar{H}_{c} = (4\pi\bar{a}^{2}/b)^{1/2}$,
so that $\xi = \epsilon \bar{\xi}$ and $\lambda = \epsilon \bar{\lambda}$.
Therefore the $\epsilon\rightarrow 0$ limit is equivalent to
taking $\xi, \lambda\rightarrow 0$ while holding $\kappa$ fixed.
In these units, the TDGL equations become (we will henceforth drop the
primes)
\begin{equation}
\epsilon^{2}(\partial_{t} + i\phi)\psi = \epsilon^{2}
\left( {\nabla\over\kappa}
  -i{\bf a}\right)^{2}\psi + \psi - |\psi|^{2} \psi +\epsilon^{3} \theta,
\label{newtdgl1}
\end{equation}
\begin{equation}
\epsilon^{2}\nabla\times\nabla\times{\bf a} = \epsilon^{2}\bar{\sigma}
\left( - {1\over \kappa}\nabla \phi - \partial_{t} {\bf a}\right)
+ {1\over 2 \kappa i} (\psi^{*} \nabla \psi - \psi\nabla\psi^{*})
  -|\psi|^{2} {\bf a} +  \epsilon^{2}\tilde{\bf J},
\label{newtdgl2}
\end{equation}
\begin{equation}
\langle \theta^{*}({\bf x},t)\theta ({\bf x}',t') = 2 \bar{T}
 \delta^{(d)}({\bf x} - {\bf x}') \delta(t-t'),
\label{newnoise1}
\end{equation}
\begin{equation}
\langle \tilde{J}_{i}({\bf x},t)\tilde{J}_{j}({\bf x}',t')\rangle =
\bar{\sigma} \bar{T} \delta^{(d)}({\bf x} - {\bf x}') \delta(t-t')
\delta_{ij}.
\label{newnoise2}
\end{equation}
To further simplify, we rewrite the order parameter in terms of an
amplitude and a phase, $\psi=f\exp(i\chi)$; in terms of the gauge-invariant
quantities ${\bf q}= {\bf a} - \nabla \chi /\kappa$ and $p=\phi
+ \partial_{t} \chi$, the magnetic and electric fields are
\begin{equation}
{\bf h} = \nabla\times {\bf q},
\label{H}
\end{equation}
\begin{equation}
{\bf e} = - {1\over \kappa} \nabla p - \partial_{t} {\bf q}.
\label{newP}
\end{equation}
Equating the real and imaginary parts of Eq.\ (\ref{newtdgl1}), and
noting that the total current has zero divergence \cite{hu72},
we arrive at the final form of the TDGL equations:
\begin{equation}
\epsilon^{2} \partial_{t} f =\epsilon^{2}\left(
{1\over \kappa^{2}} \nabla^{2} f - q^{2} f\right) + f
  - f^{3} + \epsilon^{3}\zeta,
\label{f1}
\end{equation}
\begin{equation}
\epsilon^{2}\nabla\times\nabla\times{\bf q} = \epsilon^{2}\bar{\sigma}
\left( -{1\over \kappa}\nabla p - \partial_{t}{\bf q} \right)
    -f^{2} {\bf q} + \epsilon^{2}\tilde{\bf J},
\label{Q}
\end{equation}
\begin{equation}
\epsilon^{2}{\bar{\sigma}\over\kappa} \nabla\cdot\left( {1\over \kappa}
\nabla p  + \partial_{t}{\bf q} \right) -  f^{2} p
= {\epsilon^{2}\over \kappa}\nabla\cdot \tilde{\bf J}
  -{\epsilon\over\kappa} \zeta ,
\label{P}
\end{equation}
where the noise term $\zeta$ has zero mean and correlations
\begin{equation}
\langle \zeta({\bf x},t)\zeta({\bf x}',t'\rangle) =  \bar{T}
 \delta({\bf x} - {\bf x}') \delta( t- t'),
\label{zeta}
\end{equation}
with the correlations of the current noise $\tilde{\bf J}$ given
by Eq.\ (\ref{newnoise2}) above.   Note that in these scaled units there
are three dimensionless parameters: the Ginzburg-Landau parameter $\kappa$,
which measures the coupling between the order parameter and the gauge
field, the dimensionless normal state conductivity $\bar{\sigma}$, which
determines the rate at which flux diffuses in the normal state, and the
dimensionless temperature $\bar{T}$, which measures the relative strength
of thermal fluctuations.  The remainder of this paper is
concerned with solving  Eqs.\ (\ref{f1}), (\ref{Q}),
and (\ref{P}) for a moving interface.

\section{The sharp-interface limit}

In this section we will derive a sharp interface equation from the
TDGL equations, in the limit that $\epsilon\rightarrow 0$ while
$\kappa$, $\bar{\sigma}$, and $\bar{T}$ are held fixed.  In order to
simplify the analysis we will assume that all quantities are
translationally invariant
along the direction of the applied magnetic field; in particular, the
magnetic field is ${\bf h}({\bf x},t) = h({\bf x},t) \hat{\bf z}$,
with ${\bf x}=(x,y)$.
Extending the
results to the more general three dimensional situation significantly
complicates the calculation without offering any new physical insights.
The derivation in this section is in the same spirit as derivations of
the sharp interface equations of solidification from the ``phase field
equations'' in certain distinguished limits
\cite{caginalp89,wheeler92,kupferman92}. The idea is that far from the
interface the magnetic field in the normal phase satisfies a diffusion
equation; this is the
``outer region.''  Near the interface, in the ``inner region,'' we solve
the full nonlinear TDGL equations perturbatively in the velocity and
curvature of the interface.  By matching the solutions in an
appropriate overlap region, we will see that the inner solution
provides the boundary conditions for the outer region.  This allows us
to effectively ``integrate out'' the order parameter field in favor of
a diffusion equation for the magnetic field with nonlinear boundary
conditions.  While this considerably reduces the complexity of the
problem, the remaining ``modified Stefan problem'' is very challenging
in its own right.  There has been considerable progress in recent years
in understanding the properties of this class of models within the
context of dendritic growth; see Refs. \cite{kessler88,langer88} for
reviews.

\subsection{The outer solution}

In the outer region we assume an expansion of the form
\begin{equation}
f({\bf x},t; \epsilon) = f_{0}({\bf x},t) + \epsilon f_{1}({\bf x},t)
 +\epsilon^{2} f_{2}({\bf x},t)  + \ldots,
\label{expand1}
\end{equation}
\begin{equation}
{\bf q} ({\bf x},t;\epsilon) = {\bf q}_{0}({\bf x},t)
   + \epsilon {\bf q}_{1}({\bf x},t) + \epsilon^{2} {\bf q}_{2}({\bf x},t)
    + \ldots,
\label{expand2}
\end{equation}
\begin{equation}
p({\bf x},t;\epsilon) = p_{0}({\bf x},t) + \epsilon p_{1}({\bf x},t)
 + \epsilon^{2} p_{2}({\bf x},t) + \ldots\ .
\label{expand3}
\end{equation}
Substituting these expansions into Eqs.\ (\ref{f1}),
(\ref{Q}), and (\ref{P}), and equating terms of $O(1)$,
$O(\epsilon)$, and $O(\epsilon^{2})$, we obtain the following two sets of
solutions:

{\it Superconducting solution}.  This solution corresponds to $f_{0}=1$,
$f_{1}=f_{2}=0$, ${\bf q}_{0}={\bf q}_{1}=0$, ${\bf q}_{2} = \tilde{\bf J}$,
$p_{0}=0$, $p_{1} = - \zeta/\kappa$,
$p_{2} = - \nabla\cdot\tilde{\bf J}/\kappa$.

{\it Normal solution}. This solution corresponds to $f_{0}=f_{1}=f_{2}=0$,
\begin{equation}
\nabla\times\nabla\times {\bf q}_{0} = \bar{\sigma} \left( -{1\over \kappa}
 \nabla p_{0} - \partial_{t} {\bf q}_{0} \right) + \tilde{\bf J},
\label{field1}
\end{equation}
\begin{equation}
\nabla\cdot\left[\bar{\sigma}\left(- {1\over\kappa} \nabla p_{0}
  - \partial_{t} {\bf q}_{0} \right) + \tilde{\bf J} \right] = 0 ,
\label{scalar1}
\end{equation}
with ${\bf q}_{1}$, ${\bf q}_{2}$, $p_{1}$, $p_{2}$ undetermined (note that
the second equation can be obtained by taking the divergence of the first
equation).
Taking the curl of Eq.\ (\ref{field1}) and using
$\nabla\cdot {\bf h}_{0}=0$,
we see that  the magnetic field in the normal phase satisfies the
diffusion equation
\begin{equation}
\bar{\sigma} \partial_{t} {\bf h}_{0} = \nabla^{2} {\bf h}_{0}
  + \nabla\times\tilde{\bf J}.
\label{field21}
\end{equation}
This expansion may be carried to higher order in $\epsilon$; order by order,
the magnetic field satisfies the diffusion equation (without the noise
term).  Therefore, we find that quite generally the magnetic field
in the normal phase satisfies
\begin{equation}
\bar{\sigma} \partial_{t} {\bf h} = \nabla^{2} {\bf h}
  + \nabla\times\tilde{\bf J},
\label{field2}
\end{equation}
the corrections to which are exponentially small.

\subsection{The inner solution}

To solve the TDGL equations in the inner region, we first afix a set of
local coordinates $(r,s)$ to the interface, where $r$ measures the distance
from the interface and $s$ measures the arclength along the interface.
In this coordinate system, the Laplacian becomes
\begin{equation}
\nabla^{2} = {\partial^{2} \over \partial r^{2}} + {\cal K} {\partial \over
\partial r} + (\nabla s)^{2} {\partial^{2} \over \partial s^{2}}
+ \nabla^{2} s {\partial \over \partial s},
\label{local}
\end{equation}
where ${\cal K}$ is the curvature of the interface \cite{caginalp89}.
Since the coordinates now evolve in time, the time derivatives
become
\begin{equation}
\partial_{t} \rightarrow \partial_{t} + \dot{r} {\partial \over\partial r}
 + \dot{s} {\partial \over \partial s}.
\label{time}
\end{equation}
We now ``stretch out'' the dimension normal to the interface by introducing
the scaled variable $R=r/\epsilon$.  Then, keeping the lowest order terms,
we have
\begin{equation}
\nabla^{2} = {1\over\epsilon^{2}}{\partial^{2}\over \partial R^{2}}
 + {1\over \epsilon} {\cal K} {\partial \over \partial R} + O(1),
\label{local2}
\end{equation}
\begin{equation}
\partial_{t} = - {1\over \epsilon} v_{n} {\partial \over \partial R}
+ O(1),
\label{local3}
\end{equation}
where $v_{n} = -\dot{r}$ is the velocity normal to the interface.
To ensure gauge invariance it is necessary to rescale
the vector potential so that ${\bf Q}= {\bf q}/\epsilon$.
The vector potential will be parallel to the interface, so that
${\bf Q} = Q(R,s,t) \hat{\bf t}$, with $\hat{\bf t}$ the unit vector tangent
to the interface.  Finally, since we want to treat the noise terms as first
order perturbations about the equilibrium solution, we will introduce
rescaled noise terms of the form
\begin{equation}
\bar{\zeta}(R,s,t) = \epsilon^{2}\zeta({\bf x},t), \qquad
 \bar{\bf J}(R,s,t) = \epsilon \tilde{\bf J}({\bf x},t),
\label{local31}
\end{equation}
with correlations
\begin{equation}
\langle \bar{\zeta}(R,s,t)\bar{\zeta}(R',s',t')\rangle =
 \epsilon^{3} \bar{T} \delta(R-R')\delta(s-s')\delta(t-t'),
\label{local32}
\end{equation}
\begin{equation}
\langle \bar{J}_{i}(R,s,t) \bar{J}_{j}(R',s',t')\rangle = \epsilon^{2}
\bar{\sigma} \bar{T} \delta(R-R') \delta(s-s') \delta(t-t') \delta_{ij}.
\label{local33}
\end{equation}
Then in terms of these scaled variables, the order parameter
amplitude $F(R,s,t;\epsilon) = f(r,s,t; \epsilon)$ and the
vector potential $Q(R,s,t;\epsilon) = q(r,s,t;\epsilon)/\epsilon$
satisfy
\begin{equation}
{1\over\kappa^{2}} F'' - Q^{2} F + F - F^{3} = -\epsilon \left( v_{n}
 + {1\over \kappa} {\cal K} \right) F' - \epsilon \bar{\zeta} +
O(\epsilon^{2}),
\label{local4}
\end{equation}
\begin{equation}
Q'' - F^{2} Q = - \epsilon\left( \bar{\sigma} v_{n} + {\cal K} \right) Q'
     -\epsilon \bar{J}_{t} + O(\epsilon^{2}) ,
\label{local5}
\end{equation}
where $\bar{J}_{t}=\bar{\bf J}\cdot\hat{\bf t}$ is the component of the
current noise parallel to the interface, and where the primes denote
derivatives with respect to $R$.
As before, we expand $F$, $Q$, and
$H= \nabla\times{\bf Q}=\epsilon_{ij}\partial_{i} Q_{j}$ in powers
of $\epsilon$:
\begin{equation}
F(R,s,t;\epsilon) = F_{0}(R,s,t) + \epsilon F_{1}(R,s,t) + \ldots,
\label{inner1}
\end{equation}
\begin{equation}
Q(R,s,t;\epsilon) = Q_{0}(R,s,t) + \epsilon Q_{1}(R,s,t) + \ldots,
\label{inner2}
\end{equation}
\begin{equation}
H(R,s,t;\epsilon) = H_{0}(R,s,t) + \epsilon H_{1}(R,s,t) + \ldots.
\label{inner3}
\end{equation}

Substituting these expansions into Eqs.\ (\ref{local4}) and (\ref{local5}),
at $O(1)$ we find the equilibrium Ginzburg-Landau equations,
\begin{equation}
{1\over \kappa^{2}} F_{0}'' - Q_{0}^{2} F_{0} + F_{0} - F_{0}^{3} = 0,
\label{order1}
\end{equation}
\begin{equation}
Q_{0}'' - F_{0}^{2} Q_{0}= 0,
\label{order12}
\end{equation}
with $H_{0} = Q_{0}'$.  The solutions corresponding to an interface
between the normal and superconducting phases have $F_{0}=1$, $Q_{0}=0$, and
$H_{0}=0$, for $R\rightarrow -\infty$ (the superconducting phase),
and $F_{0}=0$, $Q_{0}\sim R/\sqrt{2}$, and
$H_{0} = 1/\sqrt{2}$ for $R\rightarrow \infty$ (the normal phase).

At $O(\epsilon)$, we have
\begin{equation}
{1\over \kappa^{2}} F_{1} '' - Q_{0}^{2} F_{1} - 2 Q_{0}Q_{1}F_{0} + F_{1}
 - 3F_{0}^{2}F_{1} = - \left(  v_{n} + {1\over\kappa^{2}} {\cal K}
\right) F_{0}' - \bar{\zeta}
\label{ordere}
\end{equation}
\begin{equation}
Q_{1}'' - F_{0}^{2} Q_{1} - 2 F_{0}F_{1}Q_{0} = - \left( \bar{\sigma} v_{n}
+ {\cal K} \right) Q_{0}' - \bar{J}_{t},
\label{ordere1}
\end{equation}
\begin{equation}
H_{1} = Q_{1}' + {\cal K} Q_{0}.
\label{ordere2}
\end{equation}
The $O(\epsilon)$ perturbations satisfy a set of linear inhomogeneous
differential equations.  The homogeneous versions of these equations
have the solution $F_{1}= F_{0}'$, $Q_{1}= Q_{0}'$, which is easily seen
by direct substitution (this is a consequence of the translational
invariance of the equilibrium Ginzburg-Landau equations).  This
allows us to determine the value of $Q_{1}$ far from the interface
without explicitly solving this set of coupled equations.  To see this,
multiply Eq.\ (\ref{ordere}) by $F_{0}'$, Eq.\ (\ref{ordere1}) by
$Q_{0}'$, add, and integrate from $-\infty$ to $R$. We then integrate the
derivative terms by parts twice, using the boundary
conditions $F_{1}'(R)=F_{1}(R) = Q_{0}''(R) = 0$ for $R\rightarrow-\infty$,
and using the fact that $F_{0}'$ and $Q_{0}'$ satisfy the homogeneous
forms of Eqs.\ (\ref{ordere}) and (\ref{ordere1}).  After some
rearranging, we finally arrive at
\begin{eqnarray}
H_{0} H_{1} &+& {\cal K}Q_{0}\left({1\over \sqrt{2}} - H_{0}\right)
- Q_{1}H_{0}' + {1\over \kappa^{2}} (F_{1}'F_{0}'-F_{1}F_{0}'')\nonumber\\
 &=& - {\cal K} \int_{-\infty}^{R} dx \left[ {1\over\kappa^{2}}(F_{0}')^{2}
 + (Q_{0}')^{2} - {1\over\sqrt{2}}Q_{0}'\right] \nonumber\\
 &\ &-v_{n} \int_{-\infty}^{R} dx \left\{  (F_{0}')^{2} + \bar{\sigma}
\left[(Q_{0}')^{2} - {1\over\sqrt{2}}Q_{0}'\right] \right\} \nonumber\\
 &\ &- {1\over\sqrt{2}}\bar{\sigma} Q_{0} v_{n} - Q_{0} \bar{J}_{t}
+ \int_{-\infty}^{R} dx Q_{0} \bar{J}_{t}'
- \int_{-\infty}^{R} dx F_{0}'\bar{ \zeta} .
\label{mess}
\end{eqnarray}

\subsection{Asymptotic matching}

We are now in a position to match the inner and outer solutions.  First,
rewrite the outer expansion in terms of the inner variables, and take
$\epsilon\rightarrow 0$ while holding $R$ fixed:
\begin{eqnarray}
h(\epsilon R, s, t) &= &h_{0}(\epsilon R, s, t)
  + \epsilon h_{1}(\epsilon R, s, t) + \ldots\nonumber \\
 &=& h_{0} (0,s, t)  + \epsilon\left[ R \left.{\partial h_{0}(r,s,t) \over
 \partial r}\right|_{r=0} + h_{1}(0, s, t)\right] + O(\epsilon^{2}).
\label{match1}
\end{eqnarray}
This expansion
must match order by order onto the inner solution in the limit
$R\rightarrow \infty$.  We then have the matching conditions
\begin{equation}
\lim_{r\rightarrow \pm 0} h_{0}(r,s,t)
  = \lim_{R\rightarrow \pm \infty} H_{0}(R,s,t),
\label{match2}
\end{equation}
\begin{equation}
\lim_{r\rightarrow \pm 0}{\partial h_{0}(r,s,t)\over \partial r}
  = \lim_{R\rightarrow \pm \infty}{\partial  H_{1}(R,s,t)\over \partial R},
\label{match3}
\end{equation}
\begin{equation}
\lim_{r\rightarrow \pm 0} h_{1}(r,s,t)
     = \lim_{R\rightarrow \pm \infty}\left[ H_{1}(R,s,t)
       - R{\partial H_{1}(R,s,t)\over \partial R}\right].
 \label{match4}
\end{equation}

Using Eqs.\  (\ref{ordere1}) and (\ref{mess}), we find that at the
interface the magnetic field in the outer region is
\begin{equation}
h_{0}(0^{+},s,t) + \epsilon h_{1}(0^{+},s,t)
  = {1\over\sqrt{2}}\left[1 - \epsilon
 \left( \bar{\sigma}_{\rm ns} {\cal K} + \bar{\Gamma}^{-1} v_{n} - \bar{\eta}
\right)\right],
 \label{boundary1}
\end{equation}
where $\bar{\sigma}_{\rm ns}$ and $\bar{\Gamma}^{-1}$ are the dimensionless
surface tension and kinetic coefficient for the normal/superconducting
interface, given by
\begin{equation}
\bar{\sigma}_{\rm ns} = 2 \int_{-\infty}^{\infty} dx
\left[ {1\over\kappa^{2}} (F_{0}')^{2} + (Q_{0}')^{2}
- {1\over \sqrt{2}}Q_{0}'\right],
\label{tension}
\end{equation}
\begin{equation}
\bar{\Gamma}^{-1} = 2\bar{\sigma} \int_{-\infty}^{\infty} dx \left[
{1\over\bar{\sigma}} (F_{0}')^{2} + (Q_{0}')^{2} - {1\over \sqrt{2}}
Q_{0}' \right].
\label{friction}
\end{equation}
Some properties of the surface tension are discussed in Appendix~A,
and the kinetic coefficient is discussed in Appendix~B.
The noise term $\bar{\eta}$ is the projection of the current and order
parameter noise onto the interface,
\begin{equation}
\bar{\eta}(s,t) = - 2\int_{-\infty}^{\infty} dx\, F_{0}'(x)\,
         \bar{\zeta}(x,s,t)
  +2 \int_{-\infty}^{\infty} dx\,  Q_{0}(x)\, \bar{J}_{t}'(x,s,t).
\label{boundary2}
\end{equation}
The average of the noise is easily seen to be zero; as shown in
Appendix~C, the noise  correlations are
\begin{equation}
\langle\bar{\eta}(s,t)\bar{\eta}(s',t') \rangle =2\bar{T}\bar{\Gamma}^{-1}
 \delta(s-s')\delta(t-t'),
\label{boundary3}
\end{equation}
so that the fluctuation-dissipation theorem is satisfied \cite{hohenberg77}.
The boundary condition at the interface for the derivative of the
magnetic field in the normal phase  is
\begin{equation}
\left. {\partial h_{0}(r,s,t)\over\partial r}\right|_{r=0^{+}}
 = - {1\over \sqrt{2}} \sigma v_{n} - \bar{J}_{t}(r=0^{+},s,t).
\label{boundary4}
\end{equation}

Finally, setting $\epsilon=1$ and returning to conventional units,
we have the diffusion equation for the magnetic field,
\begin{equation}
\partial_{t} h = D \nabla^{2} h + 4\pi D \nabla\times \tilde{\bf J},
\label{boundary5}
\end{equation}
the boundary condition for the field at the interface
(denoted by the subscript $i$),
\begin{equation}
h_{i} = H_{c} \left[ 1 - {4\pi\over H_{c}^{2}}\left( \sigma_{\rm ns}
 {\cal K} + \Gamma^{-1} v_{n} - \eta \right)\right],
\label{boundary6}
\end{equation}
and a conservation condition for the field at the interface,
\begin{equation}
- \hat{\bf n}\cdot \nabla h |_{i} =D^{-1} H_{c} v_{n}
        + 4\pi \tilde{J}_{t}|_{i}.
\label{boundary7}
\end{equation}
In conventional units the surface tension and kinetic coefficient are
\begin{equation}
\sigma_{ns} = \left({H_{c}^{2}\lambda\over 4\pi}\right)
                       \bar{\sigma}_{\rm ns},\qquad
\Gamma^{-1} = \left({H_{c}^{2}\lambda\over 4\pi}\right)
 \left(2 m \gamma \over \kappa^{2}\hbar\right) \bar{\Gamma}^{-1}.
\label{boundary8}
\end{equation}

Having now derived a set of sharp interface equations from the TDGL
equations, it is instructive to compare our results with
the heuristic set of sharp interface equations,
Eqs.\ (\ref{intro1})-(\ref{intro3}), which were discussed in the
Introduction, and to compare our results with some previous work.
(1) We have chosen to focus here on the two dimensional
limit, in order to reduce the algebraic complexity of the derivation.
In three dimensions we would reinstate the vector character of $h$, and the
curvature ${\cal K}$ would be replaced by the sum of the principal
curvatures of the interface; the expressions for the surface tension and
kinetic coefficient would be unchanged.  (2) We have assumed that the
material parameters of the superconductor are isotropic, leading to an
isotropic surface tension.  One way of introducing anisotropy is to assume
that there are different effective masses along the $x$ and $y$ axes.
The analysis above  could then be easily modified along the lines
of Ref.\ \cite{caginalp86}, resulting in the more general anisotropic
boundary condition in Eq.\ (\ref{intro2}). (3) In the conservation
condition, Eq.\ (\ref{boundary7}), the critical field $H_{c}$ appears
on the right hand side multiplying $v_{n}$, whereas in  Eq.\ (\ref{intro2})
the magnetic field at the interface, ${\bf h}_{i}$, appears.  This
difference is of higher order in $\epsilon$, and has therefore been
neglected in our calculation.  (4) The boundary condition
derived above, Eq.\ (\ref{boundary6}), depends upon the interfacial
velocity, in contrast to Eq.\ (\ref{intro2}).  Such velocity dependent
corrections also arise in derivations of the sharp interface equations
of solidification from phase-field models
\cite{caginalp89,wheeler92,kupferman92}.  As discussed below, this term
will have a natural interpretation as a local viscous damping term in
the equation of motion for the interface. (5)  We have explicitly included
the effects of thermal fluctuations in deriving the sharp interface
equations, in contrast to Eqs.\ (\ref{intro1})-(\ref{intro3}), which neglect
fluctuations entirely.  In fact, our
Eqs.\ (\ref{boundary5})-(\ref{boundary7}) bear a striking resemblence
to a sharp interface model of solidification which incorporates
fluctuations, recently proposed by Karma \cite{karma93}.
(6)  Nechiporenko \cite{nechiporenko87} used the TDGL equations to study
the nucleation of the superconducting phase, and essentially derived the
inner solution in the limit of zero conductivity, and without noise.
This results in curvature driven dynamics for the interface, without
the interesting diffusion driven instabilities which occur when the
inner solution is matched onto the outer solution.

\section{The interface equation of motion}

Having now reduced the TDGL equations to a sharp interface problem,
we will carry the analysis one step further and obtain an explicit equation
of motion for the interface itself by integrating out the magnetic field.
This type of analysis has been used extensively in the study of the
solidification problem \cite{kessler88}.

\subsection{The boundary integral method}

We will consider a simply-connected superconducting domain which is
expanding into the surrounding supercooled normal phase.   The
normal/superconducting interface is a closed curve ${\cal C}$,
specified by the position vector ${\bf r}(s,t)$.  As the magnetic field in
the normal phase satisfies the diffusion equation, we start by introducing
the Green's function $G({\bf x},t|{\bf x}',t')$
for the diffusion equation,
\begin{equation}
(-D^{-1} \partial_{t'} - \nabla'^{2})G({\bf x},t|{\bf x}',t') =
  \delta^{(d)}({\bf x}- {\bf x}')\delta(t-t'),
\label{Green1}
\end{equation}
the solution of which is  (in $d$-dimensions)
\begin{equation}
G({\bf x},t|{\bf x}',t') =  D (4\pi D |t-t'|)^{-d/2}
\exp (-|{\bf x}-{\bf x}'|^{2}/4 D |t-t'|) \theta(t-t').
\label{Green2}
\end{equation}
Next, we (i) multiply Eq.\ (\ref{Green1}) by $h({\bf x},t)$, and integrate
over $({\bf x}',t')$; (ii) integrate the term involving the time derivative
of the Green's function by parts, discarding the transient term;
(iii) use the diffusion equation for $h({\bf x},t)$ to eliminate the
time derivatives.  The final result is
\begin{eqnarray}
h({\bf x},t) &= & \int_{-\infty}^{t^{+}} dt'\int_{N} d^{2}x'
\left[G({\bf x},t|{\bf x}',t') \nabla'^{2} h({\bf x}',t') - h({\bf x}',t')
\nabla'^{2}G({\bf x},t|{\bf x}',t')\right] \nonumber \\
& & \qquad +4\pi  \int_{-\infty}^{t^{+}} dt'\int_{N} d^{2}x' \,
G({\bf x},t|{\bf x}',t') \nabla'\times \tilde{\bf J}({\bf x}',t'),
\label{Green3}
\end{eqnarray}
where $t^{+}\equiv t+\delta$, with $\delta$ is arbitrarily small, and
where the subscript $N$ on the area integrals denotes an integral over the
normal phase.
Using Green's theorem, the first integral may be written as a contour
integral along the curve  ${\cal C}$; the second integral may be
integrated by parts with the final result
\begin{eqnarray}
h({\bf x},t) &= &H_{0} -  \int_{-\infty}^{t^{+}} dt'  \oint_{\cal C} ds' \,
\hat{\bf n}' \cdot [G({\bf x},t|{\bf r}',t') \nabla'h({\bf r}',t')
 - h({\bf r}',t') \nabla' G({\bf x},t|{\bf r}',t')] \nonumber \\
 & &  +4\pi \int_{-\infty}^{t^{+}} dt' \oint_{\cal C} ds'\,
\tilde{J}_{t}({\bf r}',t') G({\bf x},t|{\bf r}',t')\nonumber \\
& &  -4\pi \int_{-\infty}^{t^{+}} dt'\int_{N }
d^{2}x'\,\nabla' G({\bf x},t|{\bf x}',t')\times \tilde{\bf J}({\bf x}',t'),
\label{Green4}
\end{eqnarray}
where $H_{0}$ is the external magnetic field at the boundaries of the
sample  (the minus sign on the first integral is due to the definition
of the normal vector as pointing outward from the superconducting phase).
Using the boundary conditions at the interface,
Eqs.\ (\ref{boundary6}), (\ref{boundary7}), inside the integrals, we obtain
\begin{eqnarray}
{H_{c}\over4\pi} h({\bf x},t)& = & {H_{c} H_{0}\over 4\pi}
+ {H_{c}^{2}\over4\pi D}\int_{-\infty}^{t^{+}}dt'\oint_{\cal C} ds'\,
G({\bf x},t|{\bf r}',t') v_{n}' \nonumber \\
 & & + {H_{c}^{2}\over 4\pi}\int_{-\infty}^{t^{+}}dt'\oint_{\cal C} ds'
   \left[1 -  {4\pi \over H_{c}^{2}}\left( \sigma_{ns} {\cal K}'
     + \Gamma^{-1} v_{n}' - \eta'\right)\right]
     \hat{\bf n}'\cdot\nabla' G({\bf x},t|{\bf r}',t')\nonumber\\
  & & - H_{c} \int_{-\infty}^{t^{+}} dt' \int_{N} d^{2}x'\,
   \nabla' G({\bf x},t|{\bf x}',t') \times \tilde{\bf J}({\bf x}',t'),
   \label{Green5}
 \end{eqnarray}
where $v_{n}'\equiv v_{n}(s',t')$, etc.  This equation determines the
magnetic field in the normal phase once the shape and velocity of the
interface are specified.  This equation must also hold as ${\bf x}$
approaches the interface itself from the normal phase.  Then evaluating
Eq.\ (\ref{Green5}) on the interface ${\bf r}(s,t)$, and using the
modified Gibbs-Thomson boundary condition, Eq.\ (\ref{boundary6}),
we obtain after some rearranging
\begin{eqnarray}
\Gamma^{-1} v_{n}& +& {H_{c}^{2}\over 4\pi D}\int_{-\infty}^{t^{+}} dt'
\oint_{\cal C} ds'\, G({\bf r},t|{\bf r}',t') v_{n}'
   = {H_{c}^{2}-H_{c}H_{0}\over 4\pi} - \sigma_{ns}{\cal K}\nonumber \\
    & - & {H_{c}^{2}\over 4\pi} \int_{-\infty}^{t^{+}}dt'\oint_{\cal C} ds'
  \left[1 - {4\pi\over H_{c}^{2}}\left(\sigma_{ns}{\cal K}'
       + \Gamma^{-1} v_{n}' - \eta' \right)\right]
        \hat{\bf n}'\cdot\nabla'G({\bf r},t|{\bf r}',t')\nonumber\\
     & + & \eta +  \tilde{F} ,
\label{eqnmotion}
\end{eqnarray}
where the noise term $\tilde{F}$ is
\begin{equation}
 \tilde{F} ({\bf r},t) = H_{c} \int_{-\infty}^{t^{+}} dt' \int_{N}
 d^{2}x'\, \nabla'G({\bf r},t|{\bf x}',t')\times \tilde{\bf J}({\bf x}',t').
\label{motionnoise}
\end{equation}
The correlations of $\tilde{F}$ are discussed in Appendix~C.

Our final interface equation of motion, Eq.\ (\ref{eqnmotion}) is highly
nonlinear and nonlocal.  In order to understand the physics contained in
this equation, it is helpful to dissect it and discuss the different
terms separately.   On the left hand side, the first term provides local
viscous damping of the interface.  The second term may be viewed
as a nonlocal viscous damping term due
to the eddy currents produced in the normal phase by the moving interface.
It is this term which is responsible for the diffusive Mullins-Sekerka
instability \cite{kessler88,langer88,mullins64};  the propagation of
the superconducting phase into the normal phase is dynamically unstable
at long wavelengths.  On the right hand side, the first term
can be derived from the free energy difference between the two phases, and
is analogous to the ``undercooling'' in the solidification problem
\cite{kessler88,langer88}.   The second term arises from the surface
free energy of the interface.  The third term is a consequence of the
discontinuity of $h$ across the interface,  which results in an
effective surface current density localized
at the interface.  This produces a retarded self-interaction of the
interface;  in the limit of infinite diffusion constant (see below)
this term becomes local in time and takes the form of a Biot-Savart
interaction.
The last two terms contain the thermal fluctuations of the interface.
The equation of motion looks like a Langevin equation
with nonlocal damping and colored noise.  However, the retarded
self-interaction term on the right hand side cannot be derived from
an energy functional,  so strictly speaking our
equation is not of the Langevin form.   As shown below, the reduction to
a Langevin equation is complete in the infinite diffusion constant limit.

\subsection{The $D\rightarrow\infty$ limit}

The interface equation of motion simplifies considerably in the limit
that the diffusion constant becomes large.   In this limit the second term
on the left hand side of Eq.\ (\ref{eqnmotion}), which is responsible
for the Mullins-Sekerka instability, vanishes, as does the noise term
$\tilde{F}$. In addition, in this limit the magnetic field
satisfies Laplace's equation rather than the diffusion equation, and
the diffusion Green's functions  may be replaced
by a delta function in time multiplied by the Green's function for
Laplace's equation:
\begin{equation}
G({\bf r},t| {\bf r}',t') = - {1\over 2\pi} \ln (R/R_{0})
                \,  \delta(t-t'),
\label{largeD1}
\end{equation}
where $R=|{\bf r}(s) - {\bf r}(s')|$, and $R_{0}$ is some long
distance cutoff.  The normal gradient of $G$ is
\begin{equation}
\hat{\bf n}(s')\cdot\nabla' G({\bf r},t|{\bf r}',t') =
  {[{\bf r}(s) -{\bf r}(s')]\times \hat{\bf t}(s') \over 2\pi R^{2}}
  \,\delta(t-t').
\label{largeD11}
\end{equation}
In this limit, the interface equation of motion becomes local in time:
\begin{eqnarray}
\Gamma^{-1} v_{n}(s,t)& =& {H_{c}^{2} -H_{c} H_{0} \over 4\pi}
 - \sigma_{ns}{\cal K}(s,t) - {H_{c}^{2} \over 8\pi^{2}} \oint_{\cal C} ds'
\Biggl\{1 - {4\pi\over H_{c}^{2}}\Bigl[\sigma_{ns}{\cal K}(s',t)
 \nonumber\\
 & &\qquad  + \Gamma^{-1} v_{n}(s',t) - \eta(s',t) \Bigr]\Biggr\}
{[{\bf r}(s,t)-{\bf r}(s',t)]\times\hat{\bf t}(s',t) \over R^{2}}
+ \eta(s,t),
\label{largeD2}
\end{eqnarray}
where $\hat{\bf t}=-\hat{\bf n}\times\hat{z}$ has been used.  This last term
on the right hand side describes the self-interaction of the interface,
in the form of a Biot-Savart interaction of the current which is
flowing parallel to the interface.  In general this current is not constant,
but rather is a function of the arclength $s$.  For the moment we will
ignore this complication  and assume that the current is constant.
In addition, the in our asymptotic expansion the velocity and curvature
corrections were treated as small perturbations; this is equivalent to
assuming that the external field $H_{0}\approx H_{c}$.  With this in
mind, we see that the first term on the right hand of Eq.\ (\ref{largeD2})
is approximately $(H_{c}^{2} - H_{0}^{2})/8\pi$.  With these simplifications,
the equation of motion becomes (using
$v_{n}=\hat{\bf n}\cdot\partial_{t}{\bf r}$)
\begin{equation}
\Gamma^{-1} \hat{\bf n}\cdot\partial_{t} {\bf r}=
   {H_{c}^{2}-H_{0}^{2} \over 8\pi} - \sigma_{\rm ns} {\cal K}
    - {H_{c}^{2}\over 8\pi^{2}}\oint_{\cal C} ds' \,
    {[{\bf r}(s,t)-{\bf r}(s',t)]\times\hat{\bf t}(s') \over R^{2}} + \eta.
\label{ferro}
\end{equation}
This equation may be written in the variational form
\begin{equation}
\Gamma^{-1}\partial_{t} {\bf r} = - {1\over \sqrt{g}}
   {\delta {\cal H}_{\rm eff} \over \delta {\bf r}} +\bbox{\eta},
\label{largeD3}
\end{equation}
where $g$ is the metric for the interface (see Appendix D), and where
the effective interface Hamiltonian is
\begin{equation}
{\cal H}_{\rm eff}[{\bf r}] = - {H_{c}^{2} - H_{0}^{2} \over 8\pi}{\cal A}
 + \sigma_{ns} L + {H_{c}^{2} \over 16 \pi^{2}}
 \oint_{\cal C} ds \oint_{\cal C} ds'\, \hat{\bf t}(s)\cdot \hat{\bf t}(s')
 \ln (R/R_{0}).
\label{largeD4}
\end{equation}
Here ${\cal A}$ is the area enclosed by the curve ${\cal C}$ and $L$ is the
perimeter of ${\cal C}$ (the necessary functional derivatives are
carried out in Appendix D).   The first term is the free energy difference
between the two phases, the second term is the free energy of the interface,
and the third term is the self-interaction of the interface.  As mentioned
above, this last term is a type of Biot-Savart interaction of a
current {\it sheet}, with a current per unit length of $H_{c}/4\pi$.
The integral is therefore one-half of the self inductance of the current
sheet.

The local equation of motion, Eq.\ (\ref{largeD3}), is identical to
an equation of motion for the interface of two dimensional
ferrofluid domains recently proposed by  Langer {\it et al.} \cite{langer92}
(for an overview, see \cite{dickstein93}).   As shown by these authors,
the repulsive Biot-Savart interaction tends to favor an extended interface,
resulting in the highly branched labyrinthine patterns which are observed
when a ferrofluid droplet is subjected to an applied magnetic field.
Labyrinthine structures are also observed in the intermediate state of type-I
superconductors \cite{huebener}, and the long range Biot-Savart interaction
may play an important role in understanding the development of these
patterns \cite{dickstein93}.

\section{Discussion and summary}

In summary, we have derived an equation of motion for the
normal/superconducting interface by starting from a fluctuating version
of the TDGL equations.  In the limit of infinite diffusion constant
(zero normal state conductivity), these equations become local and
can be derived variationally from an energy functional.  For finite
diffusion constant the equation of motion becomes nonlocal, and contains an
additional term responsible for the Mullins-Sekerka instability of the
moving interface.  What remains to be studied is the competition between
the self-interaction of the interface and the Mullins-Sekerka instability,
and the implications for pattern formation in the intermediate state of
type-I superconductors \cite{jackson93}.

After this work was completed, results similar to those obtained in
Sec.\ III have been reported by Chapman {\it et al.} \cite{chapman92}.
I would like to thank Dr. Weinan E for bringing this reference to my
attention.

\acknowledgments

I would like to thank R. E. Goldstein for several useful discussions
regarding this work.
This work was supported by NSF Grants DMR 89-14051 and DMR 92-23586,
and by the Alfred P. Sloan Foundation.

\appendix

\section{The surface tension }
In this Appendix we will review some properties of the surface tension.
In addition, we will calculate the first order corrections to
surface tension near the critical value of $\kappa=1/\sqrt{2}$, leading
to Eq.\ (\ref{surface18}).  This result has not appeared in the literature
previously, and is the principle new result in this Appendix.

The canonical form of the dimensionless surface tension of the
normal/superconducting interface is \cite{fetter69}
\begin{equation}
\bar{\sigma}_{\rm ns} = \int_{-\infty}^{\infty} dx \left[ - F_{0}^{2}
+ {1\over 2} F_{0}^{4} + {1\over \kappa^{2}} (F_{0}')^{2}
+ Q_{0}^{2} F_{0}^{2} + \left( Q_{0}' - {1\over\sqrt{2}}\right)^{2}\right].
\label{surface1}
\end{equation}
To show that this is equivalent to Eq.\ (\ref{tension}), we can use
the following identity:
\begin{equation}
- F_{0}^{2} + {1\over 2} F_{0}^{4} + F_{0}^{2} Q_{0}^{2}
= {1\over \kappa^{2}} (F_{0}')^{2} + (Q_{0}')^{2} - {1\over 2}.
\label{surface2}
\end{equation}
To obtain this result, multiply the first equilibrium Ginzburg-Landau
equation, Eq.\ (\ref{order1}), by $F_{0}'$, multiply the second
Ginzburg-Landau equation, Eq.\ (\ref{order12}), by $Q_{0}'$, add the
two equations, and note that the resulting equation is an exact differential;
the constant of integration is determined by the boundary conditions on
$F_{0}$ and $Q_{0}$.  Using this identity in conjunction with
Eq.\ (\ref{surface1}) produces our result, Eq.\ (\ref{tension}).

An explicit calculation of the surface
tension requires a numerical solution of the equilibrium Ginzburg-Landau
equations.  Lacking such numerical solutions, it is still possible to
make some qualitative observations about the surface tension.  The second
set of terms in Eq.\ (\ref{tension}) which involve the vector potential
produce a negative contribution to the surface tension; for sufficiently
large $\kappa$ these terms will dominate and the surface tension will
become negative. Detailed analytic arguments \cite{saintjames69})
show that for
$\kappa\ll 1/\sqrt{2}$, $\bar{\sigma}_{\rm ns}\approx 2\sqrt{2}/3\kappa$,
so that in conventional units
$\sigma_{\rm ns}\approx 1.89 \xi (H_{c}^{2}/8\pi)$, while for
$\kappa\gg 1/\sqrt{2}$, $\bar{\sigma} = 4(\sqrt{2} -1)/3$, so that in
conventional units $\sigma_{\rm ns}\approx 1.10 \lambda (H_{c}^{2}/8\pi)$.

As first shown in the landmark paper by by Ginzburg and Landau
\cite{ginzburg50}, the surface tension vanishes at $\kappa=1/\sqrt{2}$.
Here we will extend these results somewhat, by expanding the surface
tension about the $\kappa=1/\sqrt{2}$ limit.  We will introduce a
small parameter $\tilde{\epsilon}$ such that
$1/(2\kappa^{2}) = 1+\tilde{\epsilon}$.  We then expand the equilibrium
order parameter $F_{0}$ and vector potential $Q_{0}$ in powers
of $\tilde{\epsilon}$,
\begin{equation}
F_{0}(x;\tilde{\epsilon}) = F_{00}(x) + \tilde{\epsilon} F_{01}(x) + \ldots,
\label{surface3}
\end{equation}
\begin{equation}
Q_{0}(x;\tilde{\epsilon}) = Q_{00}(x) + \tilde{\epsilon} Q_{01}(x) + \ldots,
\label{surface4}
\end{equation}
where $F_{01}$ and $Q_{01}$ and all of their derivatives vanish at
$\pm \infty$. Substituting these expansions into the surface tension,
and noting that the $O(1)$ term vanishes, we obtain
\begin{equation}
\bar{\sigma}_{\rm ns} =2 \tilde{\epsilon}  \int_{-\infty}^{\infty}
dx \left[ 2 (F_{00}')^{2} + 4 F_{00}F_{01} + 2 Q_{00}' Q_{01}'\right].
\label{surface5}
\end{equation}
Below we will solve the equilibrium Ginzburg-Landau equations for
$F_{00}'$, and derive an identity for the second and third integrals
above.

The $O(1)$ terms satisfy the equilibrium Ginzburg-Landau equations with
$1/\kappa^{2} = 2$:
\begin{equation}
2 F_{00}'' - Q_{00}^{2} F_{00} + F_{00} - F_{00}^{3} = 0,
\label{surface6}
\end{equation}
\begin{equation}
Q_{00}'' - F_{00}^{2} Q_{00} = 0.
\label{surface7}
\end{equation}
Since the surface tension vanishes at $O(1)$,  we have
\cite{saintjames69,ginzburg50}
\begin{equation}
F_{00}' = -{1\over \sqrt{2}} F_{00} Q_{00}.
\label{surface8}
\end{equation}
This equation may be used to eliminate $Q_{00}$ in Eq.\ (\ref{surface6}),
so that at $\kappa=1/\sqrt{2}$ the order parameter amplitude satisfies
\begin{equation}
2 F_{00}'' - 2 {(F_{00}')^{2}\over F_{00}} + F_{00} - F_{00}^{3}=0.
\label{surface9}
\end{equation}
As $F_{00}>0$, we can introduce a new function $u(x)$ through
$F_{00}(x) = \exp [ -u(x)]$, so that $u$ is the solution to
\begin{equation}
2 u'' + e^{-2u} - 1 = 0,
\label{surface10}
\end{equation}
with $u,\ u'\rightarrow 0$ as $x\rightarrow -\infty$.  This equation may
be integrated once; using  the boundary conditions, we obtain
\begin{equation}
u' = \left( u + {1\over 2} e^{-2u} - {1\over 2} \right)^{1/2}.
\label{surface11}
\end{equation}
This equation could be integrated again to obtain $u(x)$, and therefore
$F_{00}(x)$, but this
will not be necessary for our purposes.

At $O(\tilde{\epsilon})$, we find
\begin{equation}
2 F_{01}'' - Q_{00}^{2}F_{01} - 2Q_{00}Q_{01}F_{00} + F_{01} - 3 F_{00}^{2}
 F_{01} = - 2 F_{00}'',
\label{surface12}
\end{equation}
\begin{equation}
Q_{01}'' - F_{00}^{2}Q_{01} - 2 F_{00} F_{01} Q_{00} = 0.
\label{surface13}
\end{equation}
As in Sec. III.B, we see that the $O(\tilde{\epsilon})$ perturbations
satisfy a set of linear, inhomogeneous differential equations.  The
method of solution is identical: we multiply Eq.\ (\ref{surface12})
by $F_{00}'$, Eq.\ (\ref{surface13}) by $Q_{00}'$, add, and
integrate from $-\infty$ to $x$.   We then integrate the derivative
terms by parts twice,  and use the fact that $F_{00}'$ and
$Q_{00}'$ solve the homogeneous versions of Eqs. (\ref{surface12})
and (\ref{surface13}).  The final result is
\begin{equation}
2(F_{00}' F_{01}' - F_{00}''F_{01}) + Q_{00}'Q_{01}' - Q_{00}''Q_{01}
  = - (F_{00}')^{2}.
\label{surface14}
\end{equation}
By substituting this result into Eq.\ (\ref{surface5}), and performing a
few integrations by parts, we find that the surface tension is
\begin{equation}
\bar{\sigma}_{\rm ns} = \tilde{\epsilon}  I_{1}
\label{surface15}
\end{equation}
where the integral $I_{1}$ is given by
\begin{equation}
I_{1} = 2 \int_{-\infty}^{\infty} dx\, (F_{00}')^{2}.
\label{surface16}
\end{equation}

Since we have now expressed the $O(\tilde{\epsilon})$ correction to
the surface tension entirely in terms of $F_{00}'$, we can
now use our results above to calculate the integral.
By changing variables to $u$, $I_{1}$ may be written as
\begin{eqnarray}
I_{1}& =& 2 \int_{-\infty}^{\infty} dx\, (u')^{2} e^{-2u}\nonumber \\
 & = & 2 \int_{0}^{\infty} du\, u' e^{-2u} \nonumber \\
 & = & 2 \int_{0}^{\infty} du \left( u + {1\over 2} e^{-2u}
     - {1\over 2}\right)^{1/2} e^{-2u}.
\label{surface17}
\end{eqnarray}
The integral is rapidly convergent, and a numerical evaluation
produces $I_{1}=0.388$.  We therefore have our final result for
the surface tension near $\kappa=1/\sqrt{2}$:
\begin{equation}
\bar{\sigma}_{\rm ns}  \approx  0.388 \left( {1\over 2\kappa^{2}} - 1\right)
\qquad ({1\over 2\kappa^{2}}\approx 1).
\label{surface18}
\end{equation}

\section{The kinetic coefficient}

As with the surface tension, an explicit evaluation of the kinetic
coefficient $\bar{\Gamma}^{-1}$
defined in Eq.\ (\ref{friction}) requires a numerical
solution of the equilibrium Ginzburg-Landau equations.  In this Appendix
we will make a few simple qualitative observations, and then evaluate the
kinetic coefficient in some limiting cases.  First, note that when
$\bar{\sigma}=\kappa^{2}$, which in conventional units
implies that $2\pi\hbar\sigma/m\gamma =1$
[see Eq.\ (\ref{dimension2})],  the kinetic coefficient is proportional
to the surface tension:
\begin{equation}
\bar{\Gamma}^{-1} = \kappa^{2}\bar{\sigma}_{ns}\qquad
                     (\bar{\sigma}=\kappa^{2}).
\label{kinetic1}
\end{equation}
This is a consequence of a type of ``duality'' in the TDGL equations which
has been previously noted in the context of vortex motion in superconductors
\cite{hu72}.  However, there is generally no simple relation between the
surface tension and the kinetic coefficient.  Second, as with the
surface tension, the terms in Eq.\ (\ref{friction}) involving the
vector potential yield a negative contribution to the kinetic
coefficient.  It is therefore possible for the kinetic coefficient to
become negative for sufficiently large conductivity $\bar{\sigma}$.
By balancing the two sets of terms in the kinetic coefficient, we see that
it will become negative when $\bar{\sigma}/\kappa\sim 1$; in conventional
units we have $(2\pi\hbar\sigma/m\gamma)\sim 1/\kappa$.

In order to calculate the kinetic coefficient in the large and small
$\kappa$ limits, we can use the results of Ref.\  \cite{saintjames69}.
In the limit of small $\kappa$ the terms involving the gradient of the order
parameter amplitude dominate, and we have
\begin{equation}
\bar{\Gamma}^{-1} = {2\sqrt{2} \over 3} \kappa \qquad (\kappa\ll 1).
\label{kinetic2}
\end{equation}
In the limit of large $\kappa$ the terms involving the magnetic field
dominate, with the result
\begin{equation}
\bar{\Gamma}^{-1} = -{4\over 3} (\sqrt{2} -1) \kappa^{2}
     \left({2\pi\hbar\over m\gamma}\right) \sigma \qquad (\kappa\gg 1).
\label{kinetic3}
\end{equation}
Finally, we can also calculate the kinetic coefficient
{\it exactly} at $\kappa=1/\sqrt{2}$
by using the results derived in Appendix A above.  The result is
\begin{equation}
\bar{\Gamma}^{-1} = 0.388 \left( 1 - {2\pi\hbar\over m \gamma} \sigma\right)
  \qquad (\kappa=1/\sqrt{2}).
  \label{kinetic4}
\end{equation}
At $\kappa=1/\sqrt{2}$, the kinetic coefficient vanishes when
$2\pi\hbar\sigma/m\gamma =1$. This is to be expected, for at this value
of the conductivity the kinetic coefficient
is proportional to the surface tension, which also vanishes at
$\kappa=1/\sqrt{2}$.

\section{Noise correlations}

In order to demonstrate that the noise term $\bar{\eta}$ defined in
Eq.\ (\ref{boundary2}) satisfies the fluctuation-dissipation, we will
first work with
\begin{equation}
\bar{\eta}(R,s,t) = - 2\int_{-\infty}^{R} dx\, F_{0}' \bar{\zeta}
  +2 \int_{-\infty}^{R} dx\,  Q_{0} \bar{J}_{t}',
\label{appb1}
\end{equation}
and take the limit as $R\rightarrow \infty$ at the end of the calculation.
The average of $\bar{\eta}$ is zero.  The two point correlation is given by
\begin{eqnarray}
\langle \bar{\eta}(R,s,t)\bar{\eta}(R',s',t')\rangle& = &
 4\int_{-\infty}^{R} dx\! \int_{-\infty}^{R'}dx'\, F_{0}'(x) F_{0}'(x')
\langle \bar{\zeta}(x,s,t)\bar{\zeta} (x',s',t')\rangle\nonumber \\
& & + 4\int_{-\infty}^{R} dx\! \int_{-\infty}^{R'}dx'\,
Q_{0}(x) Q_{0}(x'){d^{2}\over dxdx'}
\langle \bar{J}_{t}(x,s,t) \bar{J}_{t}(x',s',t')\rangle.
\label{appb2}
\end{eqnarray}
Substituting the two point correlations from Eqs.\ (\ref{local31}) and
(\ref{local32}), and integrating the second term in Eq.\ (\ref{appb2})
by parts twice, we obtain
\begin{eqnarray}
\langle \bar{\eta}(R,s,t)\bar{\eta}(R',s',t')\rangle& = &
2 \bar{T} \delta(s-s') \delta (t-t') \Biggl\{ 2\int_{-\infty}^{R} dx \,
         (F_{0}')^{2} \theta(R'-x)  \nonumber \\
 & &  + 2\bar{\sigma} \Biggl[ \int_{-\infty}^{R} dx\, (Q_{0}')^{2}
 \theta(R'-x) - Q_{0}(R')Q_{0}'(R')\theta(R-R') \nonumber \\
 & &   - Q_{0}(R)Q_{0}'(R)\theta(R'-R)
                  + Q_{0}(R) Q_{0}(R') \delta(R-R')\Biggr]\Biggr\},
\label{appb3}
\end{eqnarray}
where $\theta$ is the step function.  Now take the limit
$R,R'\rightarrow \infty$ ($R\neq R'$):
\begin{eqnarray}
\langle \bar{\eta}(s,t)\bar{\eta}(s',t')\rangle
&=& \lim_{R,R'\rightarrow\infty}
\langle\bar{\eta}(R,s,t)\bar{\eta}(R',s',t')\rangle \nonumber \\
&=&2 \bar{T} \delta(s-s') \delta (t-t') \Biggl\{ 2\int_{-\infty}^{\infty}dx\,
         (F_{0}')^{2}   \nonumber \\
& & \qquad + 2\bar{\sigma}\lim_{R\rightarrow\infty}
    \Biggl[ \int_{-\infty}^{R} dx \,(Q_{0}')^{2} - Q_{0}(R)Q_{0}'(R)\Biggr]
\Biggr\}.
\label{appb4}
\end{eqnarray}
Since $\lim_{R\rightarrow \infty} Q_{0}'(R)=1/\sqrt{2}$, we may combine the
last two terms into a single integral, and finally arrive
at Eq.\ (\ref{boundary3}), with the kinetic coefficient $\bar{\Gamma}^{-1}$
defined in Eq.\ (\ref{friction}).  We thus see that the projected noise
$\bar{\eta}$ satisfies the fluctuation-dissipation theorem.

Next, we will calculate the correlations of the noise term $\tilde{F}$
defined in Eq.\ (\ref{motionnoise}).  Again, the average of $\tilde{F}$
is zero, while the two point correlation is
\begin{eqnarray}
\langle \tilde{F}({\bf r},t)\tilde{F}({\bf r}',t')\rangle &=&
H_{c}^{2} \int_{-\infty}^{t^{+}}dt_{1}\int_{-\infty}^{t'^{+}}dt_{2}
\int_{N} d^{2}x_{1} \int_{N} d^{2}x_{2} \nonumber\\
& & \qquad\times \epsilon_{ij} \epsilon_{kl} \partial_{i}
G({\bf r},t|{\bf x}_{1},t_{1})\partial_{k}G({\bf r}',t'|{\bf x}_{2},t_{2})
\langle J_{j}({\bf x}_{1},t_{1}) J_{l}({\bf x}_{2},t_{2})\rangle\nonumber\\
&=& 2\sigma k_{B} T H_{c}^{2} \int_{-\infty}^{t^{+}} dt_{1}\int_{N}
 d^{2}x_{1} \nabla_{1} G({\bf r},t|{\bf x}_{1},t_{1})\cdot
\nabla_{1} G({\bf r}',t'|{\bf x}_{1},t_{1}),
\label{appb5}
\end{eqnarray}
where Eq.\ (\ref{noise2}) has been used for the correlations of the current
noise.  By using several vector identities, along with the heat equation,
we find that this may be written as
\begin{eqnarray}
\langle\tilde{F}({\bf r},t)\tilde{F}({\bf r}',t')\rangle &=&
{H_{c}^{2} \over 4\pi D} k_{B} T [G({\bf r},t|{\bf r}',t')
 + G({\bf r}',t'|{\bf r},t)]  \nonumber \\
 & & \qquad -{H_{c}^{2}\over 4\pi} 2k_{B}T \int_{-\infty}^{t^{+}} dt_{1}
 \oint_{\cal C} ds_{1}\, \hat{\bf n}_{1}\cdot\nabla_{1}
 [G({\bf r},t|{\bf r}_{1},t_{1}) G({\bf r}',t'|{\bf r}_{1},t_{1})].
\label{appb6}
\end{eqnarray}
The first term is what would be expected for colored noise, given the
kernel in the second term on the left hand side of the equation of motion,
Eq.\ (\ref{eqnmotion}).   The physics of the second term is unclear.
It may cancel the cross-correlation $\langle \eta \tilde{F}\rangle$,
although I have been unable to demonstrate this convincingly.

\section{Differential geometry of curves in two dimensions}

Here we will review a few facts regarding the differential geometry of
closed curves in two dimensions, much of which can be found in
Refs. \cite{langer92} and \cite{brower84}.

Let ${\bf r}(\alpha)$ trace out a closed curve ${\cal C}$, with
$0\le \alpha \le 1$.
This parameterization is connected to the arclength parameterization
$s(\alpha)$ by $ds = \sqrt{g} d\alpha$, where
$g=|\partial {\bf r}/\partial \alpha|^{2}$ is the metric for the curve.
The counterclockwise tangent vector is
$\bbox{\tau}(\alpha) = \partial {\bf r}/\partial \alpha$, and the unit tangent
is $\hat{\bf t} = \bbox{\tau}/\sqrt{g}$.  In the arclength parameterization,
$\partial \hat{\bf t}/\partial s = -{\cal K} \hat{\bf n}$, where
${\cal K}$ is the curvature and $\hat{\bf n}$ is the outward unit normal,
and $\partial \hat{\bf n}/\partial s = {\cal K}\hat{\bf t}$.
The perimeter $L$ of the curve ${\cal C}$ is a functional of
${\bf r}(\alpha)$, and is given by
\begin{eqnarray}
L[{\bf r}] &=& \oint_{\cal C} ds \nonumber\\
  & = & \int_{0}^{1} d\alpha\, \sqrt{g},
\label{difgeo1}
\end{eqnarray}
and the area ${\cal A}$ enclosed by ${\cal C}$ is
\begin{eqnarray}
 {\cal A}[{\bf r}]&=&{1\over 2}\oint_{\cal C} ds\, {\bf r}
                                   \times\hat{\bf t}\nonumber\\
         &=& {1\over 2} \int_{0}^{1} d\alpha\, {\bf r}(\alpha)
               \times\bbox{\tau}(\alpha).
\label{diffgeo2}
\end{eqnarray}
Functional derivatives of these quantities are obtained by using
the Euler-Lagrange equations in the $\alpha$-parameterization, with
the results
\begin{equation}
{\delta L \over \delta {\bf r}} = \sqrt{g} {\cal K} \hat{\bf n}, \qquad
{\delta {\cal A} \over \delta {\bf r}} = \sqrt{g} \hat{\bf n}.
\label{diffgeo3}
\end{equation}
The Biot-Savart contribution to the energy is of the general form
\begin{eqnarray}
I[{\bf r}]& =& {1\over 2} \oint_{\cal C} ds\! \oint_{\cal C} ds'\,
\hat{\bf t}(s)\cdot\hat{\bf t}(s') \Phi (R) \nonumber\\
  &=& {1\over 2} \int_{0}^{1} d\alpha \int_{0}^{1}d\alpha' \,
   \bbox{\tau}(\alpha)\cdot\bbox{\tau}(\alpha') \Phi (R),
\label{diffgeo4}
\end{eqnarray}
where $ R =| {\bf r}(s)-{\bf r}(s')|$.
The functional derivative of this quantity is
\begin{equation}
{\delta I \over \delta {\bf r}} = \sqrt{g} \hat{\bf n} \oint_{\cal C}
  ds'\, {[{\bf r}(s)-{\bf r}(s')]\times \hat{\bf t}(s')\over R} \Phi'(R),
\label{diffgeo5}
\end{equation}
where $\Phi'(R)=\partial \Phi/\partial R$.

\end{document}